 \documentclass[aps,prb,onecolumn,twoside,noshowpacs,notitlepage]{revtex4}%
 \usepackage[tbtags,sumlimits]{amsmath}
 \usepackage{amssymb}

 \usepackage{xspace}
 \newtheorem{theorem}{Theorem}
 \newtheorem{lemma}{Lemma}
 \newtheorem{definition}{Definition}

 \newcommand{\Proof}{\noindent\emph{Proof:~}}
 \newcommand{\QED}{\hfill\ensuremath{\square}~\par\medskip}

 \newcommand{\ie}{i.e., }

 \newcommand{\RefEqn}[1]{Eq.~\eqref{#1}}

 \newcommand{\IFF}{$iff$\xspace}
 \newcommand{\DefEq}{\doteq}
 
 \newcommand{\abs}[1]{\ensuremath{\left\vert#1\right\vert}}
 \newcommand{\Sum}[1]{\ensuremath{\sum_{#1}}}
 \newcommand{\set}[1]{\ensuremath{{\left\{\,#1\,\right\}}}}%
 \newcommand{\setsuch}[2]{\ensuremath{\big\{\,#1\,\big|\,#2\,\big\}}}

 \newcommand{\RefCite}[1]{Ref.~\onlinecite{#1}}
 \newcommand{\One}{\openone}

 \newcommand{\TrB}[1]{{\rm Tr}_2\bigl\{\,#1\,\bigr\}}
 \newcommand{\TrC}[1]{{\rm Tr}_3\bigl\{\,#1\,\bigr\}}

 \newcommand{\ket}[1]{\ensuremath{\mathop{\vert\,{#1}\,\rangle}}}
 \newcommand{\bra}[1]{\ensuremath{\langle\,#1\,\vert}}
 \newcommand{\braket}[2]{\ensuremath{\langle#1\,\vert\,#2\rangle}}
 \newcommand{\proj}[1]{\ensuremath{\ket{{#1}}\bra{{#1}}}}

 \newcommand{\HSn}[1]{\ensuremath{\mathcal{H}{}_{#1}}\xspace}
 \newcommand{\HA}{\HSn{1}}
 \newcommand{\HB}{\HSn{2}}
 \newcommand{\HC}{\HSn{3}}

 \newcommand{\hSn}[2]{\ensuremath{\mathcal{#1}{}_{#2}}\xspace}
 \newcommand{\hA}{\hSn{G}{1}}
 \newcommand{\hB}{\hSn{G}{2}}
 \newcommand{\nA}{\hSn{N}{1}}
 \newcommand{\nB}{\hSn{N}{2}}

 \newcommand{\Halpha}{\HSn{\alpha}\xspace}
 \newcommand{\Nalpha}{\hSn{N}{\alpha}\xspace}

 \newcommand{\bRho}{\boldsymbol{\rho}}
 \newcommand{\Sys}[1]{\ensuremath{\mathcal{#1}}\xspace}

 \newcommand{\U}[1][]{\ensuremath{\boldsymbol{\rm U}%
     {\!}^{\Sys{#1}}}\xspace}

\begin{document}
\bibliographystyle{apsrev}

 \title[Uniqueness of convex sum of products of projectors]%
       {Uniqueness of a convex sum of products of projectors}
 \author{K. A. Kirkpatrick}
 \email[E-mail: ]{kirkpatrick@physics.nmhu.edu}
 \affiliation{New Mexico Highlands University, Las Vegas, New Mexico 87701}


\maketitle

\begin{center}
\parbox{4.4in}{Relative to a given factoring of the Hilbert space, the decomposition
of an operator into a convex sum of correlated products of pairs of distinct
1-projectors, one set of projectors linearly independent, is unique. }
\end{center}\ \\


\noindent Utilizing the Tridecompositional Uniqueness Theorem of Elby and
Bub{\cite{ElbyBub94}, I establish the uniqueness, relative to a given factoring of the
Hilbert space, of a decomposition of a state operator into a convex sum of correlated
products of pairs of distinct 1-projectors,
\begin{equation}\label{E:SumOfProd}
  \bRho=\Sum{j} w_j \proj{a_j}\otimes\proj{b_j},
\end{equation}
one set of projectors linearly independent.

In the appendix, I present a slightly strengthened version and simplified proof of the
Tridecompositional Uniqueness Theorem.

For the remainder of this paper I use the notation $\ket{a_j\,b_k}$ for the direct
product $\ket{a_j}\otimes\ket{b_k}$.

\section{Preliminaries}%
\noindent All vectors are normalized.
\begin{definition}
\ket{a} and \ket{b} are \emph{collinear} \IFF
$\ket{a}=e^{i\alpha}\ket{b}$, $\alpha\in\mathcal{R}$; we denote this
$\ket{a}\parallel\ket{b}$.
\end{definition}
\begin{definition}
The set \set{\ket{a_j}} is \emph{non-collinear} \IFF no pair of the set is collinear.
\end{definition}
\begin{definition}
$\bRho$ is an operator on $\HA\otimes\HB$. The \emph{null space} of $\bRho$ on \Halpha is
$\Nalpha\DefEq\setsuch{\ket{\phi}\in\Halpha}{\,\bRho\ket{\phi}=\mathbf{0}}$
($\alpha\in\set{1,2}$).
\end{definition}
\begin{lemma}
With sets \set{\ket{a_j}\in\HSn{1}}, \set{\ket{b_j}\in\HSn{2}}, and \set{w_j>0},
$j\in\set{1..N}$, and the operator $\bRho=\sum w_j\proj{\!a_j\,b_j}$, the set
\set{\ket{a_j}} spans $\hA\DefEq(\nA)^{\scriptstyle{\bot}}$ and the set
\set{\ket{b_j}} spans $\hB\DefEq(\nB)^{\scriptstyle{\bot}}$.%
\end{lemma}
\Proof  For $\ket{\phi}\in\nA$ and any $\ket{\beta}\in\HSn{2}$,
 $\bra{\phi\,\beta}\,\bRho\ket{\phi\,\beta}=0%
      =\sum w_j\abs{\braket{a_j}{\phi}}^2\abs{\braket{b_j}{\beta}}^2$;
thus $\braket{a_j}{\phi}=0\;\forall\ket{\phi}\in\nA$, so
$\ket{a_j}\in(\nA)^{\scriptstyle{\bot}}=\hA$. If \set{\ket{a_j}} does not span \hA, there
is a vector in \hA\ orthogonal to \set{\ket{a_j}}; but any such vector is annihilated by
$\bRho$ and is thus in \hSn{N}{1}, a contradiction. \QED%
The following result appears in \RefCite{Mermin99}, in the midst of the proof of another
theorem:
\begin{lemma}
\ket{\Psi} and \ket{\Phi} are vectors in $\HA\otimes\HB$. If
$\TrB{\proj{\Psi}}=\TrB{\proj{\Phi}}$, then there exists a unitary transformation \U on
\HB\ such that $\ket{\Psi}=\big(\One\otimes\U\big)\ket{\Phi}$.
\end{lemma}

\section{The uniqueness theorem}%
\begin{theorem}%
With non-collinear sets \set{\ket{a_j}\in\HSn{1}} and \set{\ket{b_j}\in\HSn{2}},
$j\in\set{1..n}$, one set linearly independent, and with non-collinear sets
\set{\ket{A_k}\in\HSn{1}} and \set{\ket{B_k}\in\HSn{2}}, $k\in\set{1..N}$, one set
linearly independent, and with sets \mbox{\set{w_j>0} and \set{W_k>0},} if
\begin{equation*}
  \sum_{j=1}^n\!w_j\proj{a_j\,b_j}%
    =\sum_{k=1}^N  W_k\proj{A_k\,B_k},
\end{equation*}
then $N=n$, and, for all $j\in\set{1..n}$,
\begin{equation*}
  \ket{A_j}\parallel\ket{a_{\pi(j)}},\;\;\ket{B_j}\parallel\ket{b_{\pi(j)}},%
    \; \text{and}\;W_j=w_{\pi(j)},
\end{equation*}
with $\pi(\cdot)$ a permutation function on \set{1..n}.
\end{theorem}
\Proof   Call the operator $\bRho$. Apply Lemma 1, with $d_1\DefEq\text{dim}\,\hA$ and
$d_2\DefEq\text{dim}\,\hB$, and recall: A set of $m$ vectors spans a space of dimension
$d\leq m$; $d=m$ \IFF the vectors are linearly independent. Without loss of generality we
take the set \set{\ket{a_j}}, which spans $\hA$, to be linearly independent; thus
$n=d_1$. Either \set{\ket{A_k}} or \set{\ket{B_k}} must be linearly independent; in
either case, $N=n$: If \set{\ket{A_k}} is linearly independent, then $N=d_1=n$.  On the
other hand, if \set{\ket{B_k}} is linearly independent, then $N=d_2$; the $n$ vectors
\set{\ket{b_j}} must span \hB, hence $n\geq d_2=N$. Similarly, the $N$ vectors
\set{\ket{A_k}} must span \hA, hence $N\geq d_1=n$, thus $N=n$.

Introduce a third Hilbert space \HC, with $\text{dim}\,\HC\geq n$; \set{\ket{c_j}} and
\set{\ket{C_j}} are orthonormal bases of \HC. Construct the two vectors
\begin{equation*}
 \ket{\psi}=\sum_{j=1}^n\sqrt{w_j}\ket{a_j\,b_j\,c_j}\quad\text{ and }\quad
 \ket{\Psi}=\sum_{j=1}^n\sqrt{W_j}\ket{A_j\,B_j\,C_j};
\end{equation*}
clearly, $\bRho=\TrC{\proj{\psi}}=\TrC{\proj{\Psi}}$. By Lemma 2, there exists a unitary
transformation \U on \HC\ such that $\ket{\psi}=(\One\otimes\One\otimes \U)\ket{\Psi}$;
defining $\ket{D_j}\DefEq\U\ket{C_j}$, we have
\begin{equation*}\sum_{j=1}^n\sqrt{w_j}\ket{a_j\,b_j\,c_j}=%
\sum_{j=1}^n\sqrt{W_j}\ket{A_j\,B_j\,D_j},\end{equation*} to which we apply Theorem~A.
\QED

\section{Discussion}%
\subsection*{``Uniqueness'' is relative to the identification of system and apparatus}%
\noindent Elby and Bub claim that \RefEqn{E:SumOfProd} ``suffers from a version of the
basis degeneracy problem.'' For example, with $\braket{a_1}{a_2}=\braket{b_1}{b_2}=0$,
the sum-of-products expression
\begin{equation}
 \bRho=\tfrac{1}{2}\,\proj{a_1\,b_1}\,+\tfrac{1}{2}\,\proj{a_2\,b_2}
\end{equation}
(which, according to Theorem~1, is unique) is the diagonalization of a degenerate
Hermitian operator (with eigenvalues $1/2$ twice, and 0 twice). The eigenvectors may be
taken to be $\ket{a_1\,b_1}$, $\ket{a_2\,b_2}$, $\ket{a_1\,b_2}$ and $\ket{a_2\,b_1}$ ---
products of vectors taken pairwise from \HSn{1} and \HSn{2}. Because of this degeneracy,
we can rotate the eigenvectors into
$\ket{q_{1,2}}=2^{-1/2}(\,\ket{a_1\,b_1}\pm\ket{a_2\,b_2}\,)$,
$\ket{q_3}=\ket{a_1\,b_2}$, and $\ket{q_4}=\ket{a_2\,b_1}$. Then
\begin{equation}\label{E:Projq}
 \bRho=\tfrac{1}{2}\,\proj{q_1}+\tfrac{1}{2}\,\proj{q_2};
\end{equation}
``the pointer reading loses its `special' status.''

This argument is flawed --- after all, the same claim may be made against the
tridecompositional uniqueness theorem itself:
\begin{equation}\label{E:Psiq}
  \ket{\Psi}=\tfrac{1}{\sqrt{2}}\bigl(\,\ket{a_1\,b_1\,c_1}\,+\,%
                       \ket{a_2\,b_2\,c_2}\,\bigr)%
  =\tfrac{1}{\sqrt{2}}\bigl(\,\ket{q_1\,d_1}\,+\,\ket{q_2\,d_2}\,\bigr),
\end{equation}
with $\ket{d_{1,2}}=2^{-1/2}\bigl(\,\ket{c_1}\,\pm\,\ket{c_2}\,\bigr)$. \RefEqn{E:Psiq}
is no more a counterexample to the tridecompositional uniqueness theorem than
\RefEqn{E:Projq} is a counterexample to Theorem 1, and for the same reason: the
``special'' nature of a pointer basis is based on the uniqueness of the form of the
decomposition in \RefEqn{E:SumOfProd}, which in turn is based on a particular
identification of system and apparatus. One cannot speak of the ``pointer basis'' without
having settled on the ``pointer'' --- the apparatus --- thus having already specified the
factor spaces.

Only having chosen a fixed identification of the subsystems (and the associated factoring
of the space) may either of these uniqueness theorems then be applied.

\appendix
\renewcommand{\thetheorem}{A}
\setcounter{theorem}{0}
\renewcommand{\thelemma}{A}
\setcounter{lemma}{0}

\section*{Appendix. The tridecompositional uniqueness theorem}%
\noindent This version of the Tridecompositional Uniqueness Theorem\cite{ElbyBub94}
avoids two assumptions of the original: that the linearly dependent set is in the same
space in each expansion, and that the expansions each have the same number of terms. The
proof here is similar to that of \RefCite{ElbyBub94}, but is considerably shorter and,
perhaps, clearer.

\begin{definition}
\ket{\Psi} is \emph{factorable} in $\HA\otimes\HB$ \IFF there exist $\ket{\alpha}\in\HA$
and $\ket{\beta}\in\HB$ such that $\ket{\Psi}=\ket{\alpha\,\beta}$.
\end{definition}
\begin{lemma}\textnormal{(Similar to Lemma 1 of \RefCite{ElbyBub94}) }
With the set \set{\ket{a_j}\in\HA} linearly independent and the set \set{\ket{b_j}\in\HB}
non-collinear,  $\ket{\Psi}=\sum_j s_j\ket{a_j\,b_j}$ is factorable in $\HA\otimes\HB$
\IFF the set \set{s_j\in\mathcal{C}} contains exactly one non-zero element.
\end{lemma}
\Proof Let $\ket{\Psi}=\ket{\alpha\,\beta}$, with $\ket{\alpha}\in\HA$ and
$\ket{\beta}\in\HB$. Expand
 $\ket{\alpha}=\sum_j a_j\ket{a_j}$, so $\ket{\Psi}=\sum_j a_j\ket{a_j\,\beta}$;
the set \set{\ket{a_j}} is linear independent, so, for each $j$,
$a_j\ket{\beta}=s_j\ket{b_j}$. For every $s_j\neq0$, $\ket{b_j}\parallel\ket{\beta}$. If
more than one $s_j\neq0$, \set{\ket{b_j}} is not non-collinear, contrary to hypothesis,
contradicting the assumption of factorability. The converse is obvious. \QED

\begin{theorem}[Tridecompositional uniqueness] With non-collinear sets
$\set{\ket{a_j}\in\HSn{1}}$, $\set{\ket{b_j}\in\HSn{2}}$, and
$\set{\ket{c_j}\in\HSn{3}}$, $j\in\set{1..n}$, two sets linearly independent, and
non-collinear sets $\set{\ket{A_k}\in\HSn{1}}$, $\set{\ket{B_k}\in\HSn{2}}$, and
$\set{\ket{C_k}\in\HSn{3}}$, $k\in\set{1..N}$, two sets linearly independent, and sets
$\setsuch{\phi_j\in\mathcal{C}}{\phi_j\neq0}$ and
$\setsuch{\varphi_k\in\mathcal{C}}{\varphi_k\neq0}$, if
\begin{equation*}
  \sum_{j=1}^n \phi_j\ket{a_j\,b_j\,c_j}
  =\sum_{k=1}^N \varphi_k\ket{A_k\,B_k\,C_k},
\end{equation*}
\noindent then $N=n$, and, for all $j\in\set{1..n}$,
\begin{equation*}
  \ket{A_j}\parallel\ket{a_{\pi(j)}},\;\;\ket{B_j}\parallel\ket{b_{\pi(j)}},\;\;%
    \ket{C_j}\parallel\ket{c_{\pi(j)}},  \text{ and }\abs{\varphi_j}=\abs{\phi_{\pi(j)}},
\end{equation*}
with $\pi(\cdot)$ a permutation function on \set{1..n}.
\end{theorem}
 \Proof    Take \set{\ket{C_k}} and \set{\ket{c_j}} to be linearly independent (with
no loss of generality: in each expansion, two of the three sets are linearly independent,
requiring coincidence in at least one space). These sets must span the same subspace of
\HC; thus $N=n$. Expand $\ket{c_j}=\sum_k\gamma_{jk}\ket{C_k}$; then
$\varphi_k\ket{A_k\,B_k}=\sum_j\phi_j\gamma_{jk}\ket{a_j\,b_j}$. For each $k$, Lemma~A
requires $\gamma_{jk}=0$ for all but one $j$; define the function
$\pi\colon\set{1..n}\rightarrow\set{1..n}$ by the relation $\gamma_{\pi(k\!)\,k}\neq0$.
We have $\ket{c_{\pi(k)}}=\gamma_{\pi(k\!)\,k}\ket{C_k}$, so
$\ket{c_{\pi(k)}}\parallel\ket{C_k}$  (and normalization requires
$\abs{\gamma_{\pi(k\!)\,k}}=1$). Because the set \set{\ket{C_k}} is non-collinear,
$\pi(\cdot)$ must be $1:1$, \ie a permutation function on \set{1..n}. We also have
$\varphi_k\ket{A_k\,B_k}=\phi_{\pi(k)}\gamma_{\pi(k\!)\,k}\ket{a_{\pi(k)}\,b_{\pi(k)}}$,
so $\ket{a_{\pi(k)}}\parallel\ket{A_k}$ and $\ket{b_{\pi(k)}}\parallel\ket{B_k}$;
normalization requires $\abs{\varphi_k}=\abs{\phi_{\pi(k)}}$. \QED


\end{document}